\documentclass[%
` aps,
  reprint,
 superscriptaddress,
bibnotes,
 amsmath,amssymb,
]{revtex4-2}
\usepackage{graphicx}
\usepackage{physics}
\usepackage{listings}

\usepackage{hyperref}
\usepackage{color}
\newcommand{\smallTimeStep}{\tau}
\newcommand{\operatorProduct}{\cdot}
\newcommand{\VEC}[1]{{\mbox{\boldmath${#1}$}}}

\usepackage[normalem]{ulem}
\makeatletter
\newsavebox{\@brx}
\newcommand{\llangle}[1][]{\savebox{\@brx}{\(\m@th{#1\langle}\)}%
  \mathopen{\copy\@brx\kern-0.5\wd\@brx\usebox{\@brx}}}
\newcommand{\rrangle}[1][]{\savebox{\@brx}{\(\m@th{#1\rangle}\)}%
  \mathclose{\copy\@brx\kern-0.5\wd\@brx\usebox{\@brx}}}
\makeatother

\usepackage{orcidlink}
\usepackage{xspace}
\usepackage{forloop,ifthen} 

\usepackage{xifthen}
\newcommand{\refAppendix}[6]{#1
  \ifthenelse{\isempty{#2}}%
    {}
    {\protect\cite{#2}}
    #3\protect\ref{#4}#5#6\xspace
}

\begin{document}

\title{ Symplectic Split-Operator Propagators from Tridiagonalized\\Multi-Mode Bosonic Hilbert Spaces for Bose-Hubbard Hamiltonians}

\author{Denys I. Bondar\orcidlink{0000-0002-3626-4804}}
\email{dbondar@tulane.edu}
\affiliation{Tulane University, New Orleans, LA 70118, USA}

\author{Ole Steuernagel\orcidlink{0000-0001-6089-7022}}
\email{Ole.Steuernagel@gmail.com}
\affiliation{Institute of Photonics Technologies, National Tsing Hua University, Hsinchu 30013, Taiwan}

\date{\today}

\begin{abstract}
  In this methods paper, we show how to tridia\-go\-nalize two families of bosonic multimode
  systems: opto\-mecha\-nical and Bose-Hubbard hamiltonians. Using tools from number theory, 
  we devise a rendering of these systems
  in the form of exact $D \times D$ tri\-dia\-go\-nal symmetric matrices with real-valued
  entries. Such matrices can subsequently be exactly dia\-go\-nalized using specialized
  sparse-matrix algorithms that need on the order of $D \ln(D)$ steps. This makes it possible to
  describe systems with much larger numbers of basis states than available to date.  It also allows
  for efficient dia\-gonal representation of large, accurate, symplectic split-operator propagators for
  which we moreover show that the required basis changes can be implemented by simple re-indexing,
  at marginal computational cost.
\end{abstract}

\maketitle

\section{Motivation\label{sec:Motivation}}

Multiparticle quantum systems occupying several coupled modes feature large numbers of basis states
which makes it hard to rigorously study them in general unless analytical solutions are known. 
No general approach for such numerically exact dia\-go\-nali\-zation yielding large numbers of
states has been found.

Here, we explicitly show, for two important quantum many-body systems, the
multi-mode opto\-mecha\-nical and the multi-site
Bose-Hubbard hamiltonian, how to render them, from the outset, in the form of numerically exact
tri\-dia\-go\-nal symmetric $D \times D$ matrices.

Subsequent exact numerical dia\-go\-nali\-zation is then possible using on the order of
${\cal{O}}(D \ln(D))$ steps~\cite{Coakley_DlnD_12,Diagonalize3DiagFootnote} instead of ${\cal{O}}(D^3)$ steps needed for
general $D \times D$ matrices, making previously intractable numbers of states accessible.

In Sections~\ref{sec:WhyDiagonalize}-\ref{sec:BHHamiltonian} we consider system structures and
symmetries to prepare for our main contribution, in Sect.~\Ref{sec:LexBases}, where we introduce
mappings achieving lexico\-graphical orderings of Fock-bases.  This is followed by applications to
Bose-Hubbard systems in Sect.~\ref{subsec:BHtridiagonalGlobal} and reports on numerical performance
in Sect.~\ref{sec:NumericalSpeeds}. 

\section{Why dia\-gonalize?\label{sec:WhyDiagonalize}}

Hamiltonians typically consistent of sums of non-commuting terms. For example in quantum-mechanical
hamiltonians $\hat {H}_{\rm qm} = \hat T + \hat V$ the terms for kinetic,
$\hat T(\hat p) = {\hat p}^2 / (2\, m)$, and potential energy,~$\hat V(\hat x)$, do not commute and,
so, we cannot perform the time propa\-gation for $\hat {H}_{\rm qm}$ in one large time-step~${t}$. The
reason lies in the infinite power series of the time-evolution opera\-tor
$\exp[- {\rm i} {t} \hat {H}_{\rm qm}/\hbar]$. Its high-order expansion features pro\-ducts of
$\hat T$ and $\hat V$~\cite{Yoshida__Proc93,Chin_PRE07,Ciric_EJPP23} which are not known in general.

In recent decades it has become clear that exponential split opera\-tor techniques provide suitable
approaches for this type of problem. Exponential symplectic split opera\-tor approaches typically conserve the
norm at machine-precision, hold down noise from the formation of derivatives, respect the symplectic
layout of phase space, are versatile~\cite{Cabrera_PRA15} and can approximate exact evolution to
high order~\cite{Javanainen_JPA06}. That is why we concentrate on them here.

Such propa\-gators use small time-steps ${ \smallTimeStep}$ and alternate between the different terms and
their respective bases.  In the case of~$\hat {H}_{\rm qm}$ we switch between position- and
momentum-basis by using fast Fourier-transforms: in these respective bases $\hat V$ and $\hat T$
are diago\-nal.

Here, we consider hamiltonians $\hat {H}_{\rm system} = \hat {H}_{\rm {0}} + \hat {H}_{\rm int}$, for
bosonic systems with { multi-site hamiltonians $\hat {H}_{\rm {0}}$ and} coupling between these sites
$\hat {H}_{\rm int}$, where $\hat {H}_{\rm {0}}$ and $\hat {H}_{\rm int}$ do not commute. With
$\hbar \equiv 1$, a one-step implementation of an exponential symplectic split-opera\-tor takes the
form
\begin{subequations}
\begin{align}
     \label{eq:ExpFullHamiltonian}
     \mathrm{Exp}[& { \smallTimeStep } \mathbf{H}_{\rm system}(t)] =  \\
     \label{eq:ExpSplitOp}
  \mathbf{S} &\cdot \mathrm{Exp}[ { \smallTimeStep } \mathbf{H}_{\rm {0}}(t) ]
               \cdot \mathbf{S}^T \cdot \mathrm{Exp}[ { \smallTimeStep } \mathbf{H}_{\rm int}(t) ] \; \cdot  \\
     \label{eq:ExpErrorTerm}
     &\cdot \mathrm{Exp}[ - {\rm i} \; {\cal O}({ \smallTimeStep}^2) ] \; ,
\end{align}                 
\end{subequations}
with a small \emph{purely complex} time step $ \smallTimeStep = - {\rm i} \Delta t $ and a second order
error term~${\cal O}({ \smallTimeStep}^2)$ that is real valued, and the propa\-gation therefore
strictly norm-preserving~\cite{Yoshida__Proc93,Cabrera_PRA15} {--to machine precision}. This holds
true even when we approximate expression~(\ref{eq:ExpFullHamiltonian}) by~(\ref{eq:ExpSplitOp}), by
dropping the error term~(\ref{eq:ExpErrorTerm}) (setting ${\cal O} = 0$).

In this work quantities given in bold font, such as the basis-changing shuffle
matrices,~$\mathbf{S}$, refer to (numerically implemented) finite square matrices.

We remind the reader that in expression~(\ref{eq:ExpSplitOp}), the exponential containing
$\mathbf{H}_{\rm int}$ can be brought to the front (or split up in `halves') yielding a third order
approximation essentially for free~\cite{Yoshida__Proc93}:
\begin{subequations}
\begin{align}
  \!\!\!     \mathrm{Exp}[& { \smallTimeStep }  \mathbf{H}_{}(t)]
           \cdot  \mathrm{Exp}[ {\rm i} \; {\cal O}({ \smallTimeStep}^3) ] =  \\
     \label{eq:ExpSplitOp3rd} 
& \mathrm{Exp}[ \tfrac{ \smallTimeStep }{2} \mathbf{H}_{\rm int}] \cdot \mathbf{S} \cdot \mathrm{Exp}[
  { \smallTimeStep } \mathbf{H}_{\rm {0}}(t) ] \cdot \mathbf{S}^T \cdot \mathrm{Exp}[ 
  \tfrac{ \smallTimeStep}{2} \mathbf{H}_{\rm int} ] 
\end{align}                 
\end{subequations}

We emphasize that, for opto\-mecha\-nical and Bose-Hubbard hamiltonians, our approach
allows us to perform basis changes via `perfect shuffle matrices', $\mathbf{S}$. Since the
$\mathbf{S}$ are unitary, they are brought `outside' of the exponential of $\mathbf{H}_{\rm {0}}$ in
expression~(\ref{eq:ExpSplitOp}); importantly, the $\mathbf{S}$ entail very low computational cost,
see Sect.~\ref{subsec:driveOMOhamiltonian}.

For the systems considered here, research has usually focused on the low temperature domain, since
in that case few states are involved. Current state-of-the-art approaches contend with using
Lanczos-type algorithms~\cite{fehske_exact_2008}. Essentially, such algorithms provide lossy
compression of the system's hamiltonian into tri\-dia\-go\-nal form such that the lowest few energy
eigenvalues and eigenfunctions can be determined `numerically exactly' (at machine
precision)~\cite{fehske_exact_2008}.

With our method, once the opera\-tors are given in dia\-gonal form for all states of interest reaching
far above the low temperature limit, their storage can be achieved at low cost and so can their
exponentiation, allowing us to study the full dynamics of the systems at high excitations.

\section{Tridiagonal rendering of tridiagonal operators\label{sec:Tri4Tri}}

Here, we are concerned with {real-valued}, finite, symmetric square matrices which contain only
three bands with nonzero entries: the dia\-go\-nal and lower and upper, immediately adjacent, sub-
and super-dia\-go\-nal bands, where these bands are symmetric in their entries $b_{n+1,n} = b_{n,n+1}$. We
call such matrices `tri\-dia\-go\-nal matrices'.

Consider the tensor pro\-duct of two matrices $\mathbf{A}$ and $\mathbf{B}$
\begin{align}\label{Eq:TensorTwoMatrics}
  \mathbf{A}\otimes\mathbf{B} = \begin{bmatrix} a_{11} \mathbf{B} & \cdots & a_{1n}\mathbf{B} \\
    \vdots & \ddots & \vdots \\ a_{m1} \mathbf{B} & \cdots & a_{mn} \mathbf{B} \end{bmatrix} .
\end{align}

Assuming matrix $\mathbf{A}$ is given in purely dia\-go\-nal form whereas $\mathbf{B}$ is tri\-dia\-go\-nal,
a moment's thought, referring to the explicit form~(\ref{Eq:TensorTwoMatrics}), reveals, that the
matrix $\mathbf{A} \otimes \mathbf{B}$ is of tri\-dia\-go\-nal form whereas
$\mathbf{B} \otimes \mathbf{A}$ is not.

The dia\-gonal matrix has to come first.

For our purposes, this is an important observation, because we want to capitalize on the fact that
tri\-dia\-go\-nal matrices can be fully dia\-go\-nalized with the necessary number of operations
scaling as low as ${\cal O}(D \ln (D))$~\cite{Coakley_DlnD_12,Diagonalize3DiagFootnote}. Here, $D$ stands
for the matrices' linear dimensions, namely,
$D_{\mathbf{A}\otimes\mathbf{B}} = D_\mathbf{A} \cdot D_\mathbf{B} $.  In general,
dia\-go\-nali\-zation of a $D \times D$ matrix is cubic in complexity, theoretically giving us an
almost quadratic speedup, namely, on the order of
${\cal O}(D^2/\ln(D))$~\cite{Coakley_DlnD_12,Diagonalize3DiagFootnote}. Moreover, to represent
symmetric real matrices only $2 D - 1$ elements need to be stored~\cite{LAPACK}.

\section{Opto\-mecha\-nical oscillator hamiltonian\label{subsec:OMOhamiltonian}}

For simplicity, let us consider a {two-mode} time-dependent opto\-mecha\-nical oscillator whose
bosonic field modes are descri\-bed by the annihilation opera\-tors $\hat a$ and $\hat b$ whilst
being driven by a classical (laser) field ${\cal E}(t) = \hbar E(t)$ such that the total system
hamiltonian (without rotating wave approximation) has the form
\begin{align}\label{Eq:Hoptomechanic}
  \hat{H}_{\rm OM}(t) = \hat{a}^\dagger \hat{a} \cdot  ( \hat{b}^\dagger + \hat{b} ) +
   {\cal E}(t)  \cdot  ( \hat{a}^\dagger + \hat{a} ) \; .
\end{align}

\subsection{Tridiagonal rendering of free opto\-mecha\-nical oscillator
  hamiltonian\label{subsec:freeOMOhamiltonian}}

First, we tri\-dia\-go\-nalize the opto\-mecha\-nical interaction between modes $a$ and $b$
\begin{align}\label{Eq:Hint}
  \hat{H}_{\rm int} = \hat{a}^\dagger \hat{a} \cdot  ( \hat{b}^\dagger + \hat{b} ) \; .
\end{align}

In the product Fock state basis, $\{\ket{n_a, n_b}\}_{n_a, n_b = 0}^{\infty}$, the bosonic number
opera\-tor $\hat{n}_a = \hat{a}^\dagger \hat{a}$ is dia\-go\-nal and the bosonic position opera\-tor
$\hat{x}_b = ( \hat{b}^\dagger + \hat{b} )/\sqrt{2} $ is tri\-dia\-go\-nal.

Using the preceding remarks of Sect.~\ref{sec:Tri4Tri}, we conclude that the matrix
$\mathbf{H}_{\rm int}$, associated with $\hat{H}_{\rm int}$, is of tri\-dia\-go\-nal form if (say on a
computer) we render it as
\begin{align}\label{Eq:TensorHint}
\mathbf{H}_{\rm int} = 
  \mathbf{n}_a \otimes \mathbf{x}_b \; .
\end{align}
E.g., the matrix for the position opera\-tor $\hat{x}_b$, is rendered as
$\mathbf{1}_a \otimes \mathbf{x}_b$ not as $\mathbf{x}_b \otimes \mathbf{1}_a$.

Consequently, the matrix $\mathbf{H}_{\rm int}$ can be efficiently dia\-go\-nalized using
${\cal O}(D \ln (D))$ steps~\cite{Coakley_DlnD_12,Diagonalize3DiagFootnote}.

Various analytical expressions for propagators of opto\-mecha\-nical oscillator hamiltonians are
known~\cite{Mancini__PRA97,Kim__PRA02}.  In Ref.~\cite{Qvarfort_NJP19} analytical solutions for
time-dependent two-mode Hamiltonians similar to~(\ref{Eq:Hoptomechanic}) were given, which from the outset
are of structurally the same tridia\-gonal
form as Eq.~(\ref{Eq:TensorHint}).

\subsection{Tridiagonal rendering of driving term of opto\-mecha\-nical 
oscillator hamiltonian\label{subsec:driveOMOhamiltonian}}

Now, in hamiltonian $ \hat{H}_{\rm OM} $~(\ref{Eq:Hoptomechanic}), consider the driving
term,~$\hat{H}_{\rm {0}}$, describing the coupling of the cavity mode $a$ to a classical laser field
${\cal E}(t) = \hbar E(t)$, namely,
\begin{align}\label{Eq:Hdrive}
  \hat{H}_{\rm {0}}(t) = {\cal E}(t)  \cdot  ( \hat{a}^\dagger + \hat{a} )  \otimes \mathbf{1}_b \; .
\end{align}
According to Sect.~\ref{sec:Tri4Tri}, the position
opera\-tor~$ ( \hat{a}^\dagger + \hat{a} )/\sqrt{2}$ is not tri\-dia\-go\-nal when rendered in
the matrix-basis $\mathbf{x}_a \otimes \mathbf{1}_b$ we used before. Instead, we have to use
$\mathbf{1}_b \otimes \mathbf{x}_a$.

Following the logic of Eq.~(\ref{Eq:TensorHint}) we need to bring matrix
$ \mathbf{H}_{\rm {0}}$ into a form compatible with the matrix-basis
$\mathbf{1}_a \otimes \mathbf{1}_b$.
We use `Perfect Shuffle Matrices' for this task.

\subsection{Perfect Shuffle Matrices  $\mathbf{S}$ \label{subsec:PerfectShuffle}}
  
Although, generally, $ \mathbf{A}\otimes\mathbf{B}$ and $\mathbf{B}\otimes\mathbf{A}$ are not equal,
they are permutation
equivalent:~$ \mathbf{A} \otimes \mathbf{B} = \mathbf{S} \cdot (\mathbf{B}\otimes\mathbf{A}) \cdot
\mathbf{S}^T$, where, as is known from the theory of Kronecker pro\-ducts, $\mathbf{S}$ are `perfect
shuffle matrices'. These are sparse with exactly {$D_{\mathbf{A}\otimes\mathbf{B}}$ nonzero
  entries}, of unit value each. The $\mathbf{S}$ permute whole rows and whole columns and invert
their own transposes $\mathbf{S}^T$.  We emphasize that the $\mathbf{S}$ can therefore be
implemented in terms of simple index permutations such that these operations have a very small
computer memory footprint.

\section{Bose-Hubbard systems\label{sec:BHHamiltonian}}

We consider a Bose-Hubbard system described by the hamiltonian $ \hat{H}_{\rm BH}$. For
transparency we first discuss only three bosonic field modes $\hat b_j, \{j=1,2,3\}$
($[\hat b_j,\hat b_k^\dagger]=\delta_{j,k} \hat{1}$), in Section~\ref{subsec:BHtridiagonalReorder}
we extend our results to $K$ sites.
\begin{subequations}
  \label{eq:BHH}
  \begin{align}\label{Eq:BHHamiltonian}
    \hat{H}_{\rm BH} =  & \sum_{j=1}^{3} \hat{d}_{j} +  \sum_{j=1}^{3} \hat{h}_{j,j+1} \\
    \label{Eq:BHHamiltonianOnsite}
    =   &  -  \sum_{j=1}^{3} \mu_j \hat n_{j} 
          +  \sum_{j=1}^{3} \tfrac{U_j}{2} \hat n_{j}(\hat n_{j} - \hat 1) \\
    \label{Eq:BHHamiltonianHop}
        &  - \sum_{j=1}^{2} J_j ( \hat b^\dagger_{j} \hat b_{j+1}  + \hat b_j \hat b^\dagger_{j+1} ) \; .
  \end{align}
\end{subequations}
Here, the dia\-gonal term~$\hat{d}_{j}$ contains the chemical potentials, $\mu_j$, and the intra-site
repulsion strengths, $U_j$, whereas the hopping term,~$\hat{h}_{j,j+1}$, contains the nearest neighbor
hopping rates,~$J_{j}$. 
Although our approach supports such site-dependent coefficients, from now on, we present $\mu$, $U$ 
and $J$ as site-independent. This allows us to give concise statements and to emphasize 
when ``code recycling'' can be applied easily.

With the number opera\-tors $\hat n_j = \hat b^\dagger_j \hat b_j$, we employ the pro\-duct Fock
space basis $\{\ket{n_1, n_2, n_3}\}_{n_1, n_2, n_3 = 0}^{\infty,\infty,\infty}$.  Whilst the matrix
renderings $\mathbf{d}_{j}$ of the 3-tuples of on-site interaction
terms~(\ref{Eq:BHHamiltonianOnsite}) are dia\-go\-nal in this basis, the 2-tuples of two inter-site
hopping terms $\mathbf{h}_{j,j+1}$ in expression~(\ref{Eq:BHHamiltonianHop}) can be given
tridia\-gonal form.

We therefore combine these expressions into the tridia\-gonal terms
$ \mathbf{H}_{j,j+1} = \mathbf{d}_{j} + \mathbf{h}_{j,j+1} $. Now we consider how to implement the
shuffle matrix basis changes,~$\hat{S}$.

\subsection{Shuffle Matrices for 3-site Bose-Hubbard system with periodic boundary conditions\label{sec:shuffleBH3}}

The hamiltonian for a 3-site Bose-Hubbard system with periodic boundary conditions has the form
\begin{align}
    \hat{H}_{BH}^{\rm [3, \textrm{periodic}]} &= \hat{H}_{1,2} + \hat{H}_{2,3} + \hat{H}_{3,1} \; ,
\label{eq:HBHintro}
\end{align}
\begin{align}
  \text{with } \;     \hat{H}_{1,2} =& -J(\hat{b}^\dagger \otimes  \hat{b} \otimes \hat{1} + \hat{b} \otimes  \hat{b}^\dagger \otimes \hat{1})
      -\mu \hat{n} \otimes \hat{1} \otimes \hat{1} \notag\\
    &+ \frac{U}{2} \hat{n} (\hat{n} - \hat{1}) \otimes \hat{1} \otimes \hat{1}, 
\end{align}
\begin{align}  
     \hat{H}_{2,3} =& -J(\hat{1} \otimes \hat{b}^\dagger \otimes  \hat{b}   + \hat{1} \otimes \hat{b} \otimes  \hat{b}^\dagger)
      -\mu \hat{1} \otimes  \hat{n} \otimes \hat{1} \notag\\
    &+ \frac{U}{2} \hat{1} \otimes \hat{n} (\hat{n} - \hat{1}) \otimes \hat{1} ,
  \\
  \text{and } \;  \hat{H}_{3,1} =& -J(\hat{b} \otimes \hat{1} \otimes \hat{b}^\dagger   + \hat{b}^\dagger  \otimes \hat{1} \otimes \hat{b}  )
      -\mu \hat{1} \otimes \hat{1} \otimes  \hat{n}  \notag\\
    &+ \frac{U}{2} \hat{1} \otimes \hat{1} \otimes \hat{n} (\hat{n} - \hat{1}) .
\label{eq:HBHintroLastTerm}
\end{align}

Since, for a periodic three site system, perfect shuffle matrices perform the roll-over operation
\begin{align} 
    \hat{S} (\hat{A} \otimes \hat{B} \otimes \hat{C}) \hat{S}^T = \hat{C} \otimes \hat{A} \otimes \hat{B}  \; ,
\label{eq:3rollOver}
\end{align}
and, for three sites, also obey the periodicity condition
\begin{align}
\!    \hat{S}^3 = 1 \quad
    \Longrightarrow \quad \hat{S}^T\hat{S}^T\hat{S}^T = 1
    \quad \Longrightarrow \quad \hat{S}^T\hat{S}^T = \hat{S} .
    \label{eq:period3}
\end{align}
We have with 
\begin{align}
    \hat{S} \hat{H}_{1,2} \hat{S}^T = \hat{H}_{2,3}, \qquad
    \hat{S} \hat{H}_{2,3} \hat{S}^T = \hat{H}_{3,1}, \notag\\
    \textrm{and } \qquad
    \hat{H}_{3,1} = \hat{S}\hat{S} \hat{H}_{1,2} \hat{S}^T \hat{S}^T \; ,
    \label{eq:3terms}
\end{align}
that
\begin{align}
    \hat{H}_{BH}^{\rm [3, \textrm{periodic}]} &= \hat{H}_{1,2} +\hat{S} \hat{H}_{1,2} \hat{S}^T + \hat{S}^T \hat{H}_{1,2} \hat{S} \; .
    \label{eq:BH_with_3terms}
\end{align}
Following Eq.~(\ref{eq:ExpSplitOp}) and using expressions~(\ref{eq:period3}),~(\ref{eq:3terms})
and~(\ref{eq:BH_with_3terms}), we generate the associated
Trotter-expansion~\cite{Hatano_LN05,Yoshida__Proc93}
\begin{align}
\label{eq:Unitary_3_Ascending}
    \overleftarrow{\mathrm{Exp}}&[{ \smallTimeStep } \hat{H}_{BH}^{\rm [3, \textrm{periodic}]}] - \mathcal{O}( \smallTimeStep^2)  \notag\\
    &= { \rm e }^{{ \smallTimeStep } \hat{S}\hat{S} \hat{H}_{1,2} \hat{S}^T \hat{S}^T} { \rm e }^{{ \smallTimeStep } \hat{S} \hat{H}_{1,2} \hat{S}^T} { \rm e }^{{ \smallTimeStep } \hat{H}_{1,2}}  \notag\\
    &= \hat{S}^T { \rm e }^{{ \smallTimeStep } \hat{H}_{1,2}}  \hat{S}^T { \rm e }^{{ \smallTimeStep } \hat{H}_{1,2}} \hat{S}^T { \rm e }^{{ \smallTimeStep } \hat{H}_{1,2}} 
   \; ,
\end{align}
where the symbol $\overleftarrow{\mathrm{Exp}}$ emphasizes the fact that this operator product,
multiplied onto a state, applies hamiltonians in ascending order, those with higher index $j$ are
applied after those with lower index.

For the descending order expression, $\overrightarrow{\mathrm{Exp}}$, we get
\begin{align}
\label{eq:Unitary_3_Descending}
    \overrightarrow{\mathrm{Exp}}&[{ \smallTimeStep } \hat{H}_{BH}^{\rm [3, \textrm{periodic}]}] - \mathcal{O}( \smallTimeStep^2)\notag \\
    &= { \rm e }^{{ \smallTimeStep } \hat{H}_{1,2}} \hat{S} { \rm e }^{{ \smallTimeStep } \hat{H}_{1,2}} \hat{S} { \rm e }^{{ \smallTimeStep } \hat{H}_{1,2}} \hat{S} \; ,
\end{align}
which conforms with expression~(\ref{eq:Unitary_3_Ascending}).

\subsection{Shuffle Matrices for the \protect{$K$}-site case with periodic boundary conditions\label{sec:shuffleBH-K}}

Consider $K$-site Bose-Hubbard systems with periodic boundary conditions, their hamiltonians,
in analogy to Eqns.~(\ref{eq:HBHintro})-(\ref{eq:HBHintroLastTerm}), have the form
\begin{align}
\label{eq:HBH_K_Periodic}
    \hat{H}_{BH}^{ [K, \textrm{periodic}]} &= \sum_{j = 1}^{K} \hat{H}_{j,\, (j + 1)\bmod K} \; .
\end{align}
When using $\hat{S}$ for $K$ sites the periodicity conditions become
\begin{align}
\label{eq:_K_PeriodicityCondition}
    \hat{S}^K = 1 \quad \textrm{  or  }\quad  
    (\hat{S}^T)^{K-1} = \hat{S} \; 
\end{align}
and the roll-over operation~(\ref{eq:3rollOver}) generalizes to
\begin{align}
    \hat{S} \left( \otimes_{j=1}^{K} A_j \right) \hat{S}^T
    = A_K \otimes_{j=1}^{K - 1} A_j \; .
\end{align}
With
\begin{align}
    & \hat{S} \hat{H}_{j,j + 1} \hat{S}^T = \hat{H}_{(j + 1) \bmod K, (j + 2)\bmod K} 
\label{eq:Coupling_H_recycle}
\end{align}
for {$j \leq K$ we} get
\begin{align}   
    & \hat{H}_{j, (j + 1)\bmod K} = \hat{S}^{j - 1} \hat{H}_{1,2} (\hat{S}^T)^{j - 1}
\end{align}
and so
\begin{align}
\label{eq:HBH_K_Periodic_shuffleForm}
\hat{H}_{BH}^{ [K, \textrm{periodic}]} &= \sum_{j = 1}^{K} \hat{S}^{j - 1} \hat{H}_{1,2} (\hat{S}^T)^{j - 1} \; .
\end{align}
or
\begin{align}
\label{eq:Unitary_K_Descending_PERIODIC_long}
    \overrightarrow{\mathrm{Exp}}&[{ \smallTimeStep } \hat{H}_{BH}^{ [K, \textrm{periodic}]}] - \mathcal{O}( \smallTimeStep^2) \notag \\
    = &\quad { \rm e }^{{ \smallTimeStep } \hat{H}_{1,2}} \hat{S} { \rm e }^{{ \smallTimeStep } \hat{H}_{1,2}} \hat{S}^T \hat{S}^2 { \rm e }^{{ \smallTimeStep } \hat{H}_{1,2}} (\hat{S}^T)^2 \notag\\
    & \qquad \operatorProduct \ldots \operatorProduct \hat{S}^{K-2} { \rm e }^{{ \smallTimeStep } \hat{H}_{1,2}} (\hat{S}^T)^{K-2}  \notag\\
    & \qquad \operatorProduct \hat{S}^{K-1} { \rm e }^{{ \smallTimeStep } \hat{H}_{1,2}} (\hat{S}^T)^{K-1} \; .
\end{align}
In summary, generalizing expression~(\ref{eq:Unitary_3_Descending}) for descending ordering, we get
\begin{align}
\label{eq:Unitary_K_Descending_PERIODIC}
    \overrightarrow{\mathrm{Exp}}&[{ \smallTimeStep } \hat{H}_{BH}^{ [K, \textrm{periodic}]}] = \left({ \rm e }^{{ \smallTimeStep } \hat{H}_{1,2}} \hat{S}\right)^K + \mathcal{O}( \smallTimeStep^2) \; ,
\end{align}
whereas expression~(\ref{eq:Unitary_3_Ascending}) for ascending ordering generalizes to
\begin{align}
\label{eq:Unitary_K_Ascending_PERIODIC}
    \overleftarrow{\mathrm{Exp}}&[{ \smallTimeStep } \hat{H}_{BH}^{ [K, \textrm{periodic}]}] = \left(\hat{S}^T { \rm e }^{{ \smallTimeStep } \hat{H}_{1,2}} \right)^K + \mathcal{O}( \smallTimeStep^2) \; ,
\end{align}

\subsection{$K$-site Bose-Hubbard system with open boundary conditions and its  Shuffle Matrices\label{sec:shuffleBHKopen}}

Choosing open boundaries at the ends of a string of $K$ sites $(\hat H_{K,1} = 0)$, hamiltonian
$ \hat{H}_{BH}^{ [K, \textrm{periodic}]}$ of Eq.~(\ref{eq:HBH_K_Periodic_shuffleForm}) gets modified to
\begin{align}
\label{eq:HBH_K_OPEN_v0}
\!\!\!     \hat{H}_{BH}^{ [K, \textrm{open}]} = & \sum_{j = 1}^{K - 1} \hat{S}^{j - 1} \hat{H}_{1,2} (\hat{S}^T)^{j - 1} 
    + \hat d_{K}    \\
  \label{eq:HBH_K_OPEN}
     = & \sum_{j = 1}^{K - 1} \hat{S}^{j - 1} \hat{H}_{1,2} (\hat{S}^T)^{j - 1} 
    +   \hat{S}^T \hat d_{1} (\hat{S}^T)^{K - 1 } ,
\end{align}
where, in expression~(\ref{eq:HBH_K_OPEN}), we used $\hat{S}^{K - 1} = \hat{S}^T$ from
Eq.~(\ref{eq:_K_PeriodicityCondition}). With Eq.~(\ref{eq:Unitary_K_Ascending_PERIODIC}), its first
order Trotter expansion~\cite{Hatano_LN05} can therefore, in ascending order, be written as
\begin{align}
\label{eq:Unitary_K_Ascending_OPEN}
    \overleftarrow{\mathrm{Exp}}&[{ \smallTimeStep } \hat{H}_{BH}^{ [K, \textrm{open}]}] - \mathcal{O}( \smallTimeStep^2)  \notag \\
    = & \; {\hat S^T}  \operatorProduct
        { \rm e }^{{ \smallTimeStep } \hat{d}_{1}} \operatorProduct \overleftarrow{\mathrm{Exp}}[{ \smallTimeStep } \hat{H}_{BH}^{ [K-1, \textrm{periodic}]}]    \; ,
\end{align}
and in descending order, as
\begin{align}
\label{eq:Unitary_K_Descending_OPEN}
  \overrightarrow{\mathrm{Exp}}&[{ \smallTimeStep } \hat{H}_{BH}^{ [K, \textrm{open}]}] 
  - \mathcal{O}( \smallTimeStep^2)  \notag \\
  = & \;  \overrightarrow{\mathrm{Exp}}[{ \smallTimeStep } \hat{H}_{BH}^{ [K-1, \textrm{periodic}]}]
      \operatorProduct   { \rm e }^{{ \smallTimeStep } \hat{d}_{1}} \operatorProduct  {\hat S}  \; .
\end{align}
Our use of the $\overrightarrow{\mathrm{Exp}}$ and $\overleftarrow{\mathrm{Exp}}$ symbols emphasizes
that this rendering allows for obvious code-recycling when implemented on a computer, see remarks
after introduction of the hamiltonian~(\ref{eq:BHH}) though.  

Note, that in
expressions~(\ref{eq:Unitary_K_Ascending_OPEN}) and~(\ref{eq:Unitary_K_Descending_OPEN}) all shuffle
matrices, including those implicit in \emph{all} of the $\overrightarrow{\mathrm{Exp}}$ and
$\overleftarrow{\mathrm{Exp}}$ symbols, are of order $K$, see
Eq.~(\ref{eq:_K_PeriodicityCondition}), \emph{not} of order $K-1$.

\subsection{Approximation of $K$-site Bose-Hubbard Propagators up to Second order in 
time step $ { \smallTimeStep } $ \label{sec:_BHK_Unitary_2ndOrder}}

The first order expansions in Sections~\ref{sec:shuffleBH3}-\ref{sec:shuffleBHKopen} can be converted
into second order expansions~(\ref{eq:ExpSplitOp3rd}) using the Baker-Hausdorff
decomposition~\cite{Hatano_LN05,Yoshida__Proc93}

\begin{align}
\label{eq:BakerHausdorff_2nd}
   {\rm  e}^{[ \smallTimeStep (\hat{A} + \hat{B})]} - \mathcal{O}( \smallTimeStep^3)  
    = 
       {\rm  e}^{[ \frac{\smallTimeStep}{2} \hat{A}]}  \; {\rm  e}^{[ \smallTimeStep \hat{B}]}  \; {\rm  e}^{[ \frac{\smallTimeStep}{2} \hat{A}]} 
      \; .
\end{align}

$K$-fold repeated application of relation~(\ref{eq:BakerHausdorff_2nd}) to hamiltonian~(\ref{eq:HBH_K_Periodic_shuffleForm}) {[here, formally $\hat B = \hat{1}$]}
yields
\begin{align}
   {\mathrm{Exp}}&[{ \smallTimeStep } \hat{H}_{BH}^{ [K, \textrm{periodic}]}] - \mathcal{O}( \smallTimeStep^3)  
   \notag \\
   \label{eq:Unitary_K_PERIODIC_2ndOrder_shuffleForm}
   = \; &    \left({ \rm e }^{ { \tfrac{\smallTimeStep}{2}} \hat{H}_{1,2}} \hat{S}\right)^K \operatorProduct \left(\hat{S}^T { \rm e }^{ {\tfrac{\smallTimeStep}{2}} \hat{H}_{1,2}} \right)^K  \\
      \label{eq:Unitary_K_PERIODIC_2ndOrder}
   = \; & \overrightarrow{\mathrm{Exp}}[\tfrac{\smallTimeStep}{2} \hat{H}_{BH}^{ [K, \textrm{periodic}]}] \operatorProduct \overleftarrow{\mathrm{Exp}}[\tfrac{\smallTimeStep}{2} \hat{H}_{BH}^{ [K, \textrm{periodic}]}] \; .
\end{align}
In the case of an open boundary between site $K$ and site~{`1'} we get, similarly, after $(K-1)$-fold
repeated application of relation~(\ref{eq:BakerHausdorff_2nd}) to the
hamiltonian~(\ref{eq:HBH_K_OPEN}), 
\begin{align}
   {\mathrm{Exp}}&[{ \smallTimeStep } \hat{H}_{BH}^{ [K, \textrm{open}]}] - \mathcal{O}( \smallTimeStep^3)  
   \notag \\
  \label{eq:Unitary_K_OPEN_2ndOrder_shuffleForm}
   = \; &    \left({\rm e}^{ \frac{\smallTimeStep}{2} \hat{H}_{1,2}} \hat{S}\right)^{K-1} \operatorProduct {\rm e}^{{ \smallTimeStep } \hat{d}_{1}} 
          \operatorProduct \left(\hat{S}^T {\rm e}^{\frac{\smallTimeStep}{2} \hat{H}_{1,2}} \right)^{K-1}  \\
  = \; & \overrightarrow{\mathrm{Exp}}[\frac{\smallTimeStep}{2} \hat{H}_{BH}^{ [K-1, \textrm{periodic}]}] 
  \operatorProduct {\rm e}^{{ \smallTimeStep } \hat{d}_{1}}
    \operatorProduct 
\overleftarrow{\mathrm{Exp}}[\frac{\smallTimeStep}{2} \hat{H}_{BH}^{ [K-1, \textrm{periodic}]}] \notag \\ 
\label{eq:Unitary_K_OPEN_2ndOrder}
   = \; & \overrightarrow{\mathrm{Exp}}[\frac{\smallTimeStep}{2} \hat{H}_{BH}^{ [K, \textrm{open}]}] 
    \operatorProduct 
\overleftarrow{\mathrm{Exp}}[\frac{\smallTimeStep}{2} \hat{H}_{BH}^{ [K, \textrm{open}]}] \; ,
\end{align}
which agrees with expressions~(\ref{eq:Unitary_K_Ascending_OPEN})
and~(\ref{eq:Unitary_K_Descending_OPEN}).

We emphasize that, because our approach uses unitary propagators, such second order (and potentially
higher order~\cite{Hatano_LN05,Yoshida__Proc93}) approximants can readily be implemented.

\section{Lexicographically ordered bases\label{sec:LexBases}}

\subsection{Preliminary considerations for local tri\-dia\-go\-nal
  rendering\label{subsec:BHtridiagonalPreliminary}}

For definiteness, let us concentrate on the hopping-term $\mathbf{H}_{1,2}$ of
hamiltonian~(\ref{eq:BHH}) and let us single out a dia\-go\-nal matrix entry, say, the one for
$\ket{n_1, n_2,n_3}$. For the next few paragraphs, we can assume that the third mode is unaffected
and we therefore drop reference to the third entry~$n_3$ entirely. With this provision, we notice that,
according to Eq.~(\ref{Eq:BHHamiltonianOnsite}) the entry~ $\ket{n_1, n_2}$ of $\mathbf{d}_{1}$ maps
onto itself with strength $-\mu n_1 + \frac{U}{2} n_1(n_1-1) $.

We desire to simultaneously enforce the following tridia\-gonal form for the off-dia\-go\-nal hopping term
$\mathbf{H}_{1,2}$ of~(\ref{Eq:BHHamiltonianHop}): the sub-dia\-go\-nal term (in the
matrix-rendering, this is the term immediately \emph{below} the entry for $\ket{n_1, n_2}$), maps
from $\ket{n_1, n_2}$ to $\ket{n_1+1, n_2-1}$ [with strength~$J_1 \sqrt{(n_1+1)n_2}$] whilst the
adjacent super-dia\-go\-nal term, the one immediately \emph{to the right}, maps $\ket{n_1+1, n_2-1}$
to $\ket{n_1, n_2}$ [\emph{also} with strength~$J_1 \sqrt{(n_1+1)n_2}$].

We achieve this goal globally by lexicographically ordering the Fock state product basis such that locally,
on the dia\-go\-nal, the entry for $\ket{n_1, n_2}$ is imme\-dia\-tely preceded by
$\ket{n_1-1, n_2+1}$ and imme\-dia\-tely succeeded by $\ket{n_1+1, n_2-1}$.  This preserves the
dia\-gonal form of $\mathbf{H}_{j,j}$ and simultaneously results in a tridia\-gonal form for the
matrix~$\mathbf{H}_{1,2}$.

Reintroducing the third entry~$n_3$ back into our discussion, we can also state the above
considerations as follows: we want to construct a mapping of the Fock pro\-duct basis onto
entries along the dia\-go\-nal which fulfil the recurrence relation
\begin{align}
\label{eq:recurrence}
  z(n_1 \pm 1, n_2 \mp 1, n_3) = z(n_1, n_2, n_3) \pm 1\; .
\end{align}
Here, $z$ stands for the non-negative integer index that enumerates the dia\-go\-nal of lexicographically ordered
dia\-go\-nal matrices (such as $\mathbf{n}_{j}$), their assoc\-ia\-ted eigenvectors we call
$\| z \rrangle$.

For global tri\-dia\-go\-nal rendering of matrix $\mathbf{H}_{1,2}$ we need three main
ingredients: a mapping from the 3-tuple `pro\-duct' basis $\{\ket{n_1, n_2, n_3}\}$ to
$\mathbf{H}$'s `linearly' arranged lexicographically ordered basis $\{\| z \rrangle\}$ which is
invertible, fulfilling the recurrence relations~(\ref{eq:recurrence}), and leaving no gaps in $z$.

Invertibility is needed to switch solutions back to the original pro\-duct Fock space.

The no-gaps requirement for $z=0,1,2,...,D$ is needed because the existence of gaps, if nonzero
terms for hopping across gaps exist, would imply failure since it gives rise to terms in
$\mathbf{H}$ which are at least as far removed from the dia\-go\-nal as the gap-width, breaking
tridiagonality.

\subsection{Tridiagonalization with lexicographically ordered bases\label{subsec:BHtridiagonalReorder}}

The requirements for mappings from pro\-duct to line\-ar bases, just laid out in
Sect.~\ref{subsec:BHtridiagonalPreliminary}, can be met using `Combinatorial Number Systems', also
known as `Macaulay representations', of the non-negative integers~$\mathbb{N}_0$ (zero is included with the other
positive integers).

Moreover, the absence of gaps allows us to render $\mathbf{H}$ using $2 D-1$
terms, where the linear dimension of the matrix $D$ is ideally given by the pro\-duct of maximum
excitation of the $K$ modes $N_j$ used in a computation: namely $D = \prod_{j=1}^{K} N_j$, but no
larger value.

If we can rely on the conservation of total excitation number $N = \sum_{j=1}^K N_j$ across $K$
sites and if we, additionally, chose an initial quantum state exclusively formed from a
superposition of basis states from the same occupation number `island' of states with one fixed
value $N$ only, then, this island's much smaller total size, ${\cal N}_N^K $, can be chosen for
the dimensions~$D$ of the matrices, see discussion around expression~(\ref{eq:NumberInIsland})
below.

The lexicographical ordering of the basis therefore has to respect the ``excitation-island
property'' of rendering each island `continuously', without gap, as well as sequen\-tial gap-less
position\-ing of islands.

We now show how to generate such lexicographically ordered computational bases
$\{\|{z}\rrangle\}_{z = 0}^{D}$ from $K$-tuple states of general $K$ mode pro\-duct Fock spaces
$\{\ket{n_1, n_2, \ldots , n_K}\}_{n_1, n_2, \ldots , n_K = 0}^{N_1, N_2, \ldots , N_K}$,
{subject to the excitation islands' number constraints  $N = \sum_{j=1}^K n_j$.}

\subsection{Combinatorial Number Systems\label{subsubsec:CombNumberSystem}}

In a Combinatorial Number System, any integer $V$ is represented by a $K$-combination, which are
constructed by choosing and fixing the positive integer $K$ and then apply\-ing the following
`greedy' algorithm: find the largest value $v_K$~\cite{BisectionFootnote} such that the binomial
coefficient obeys $\binom{v_K}K \leq V$ and then subtract it from $V$. Lower $K$ by one and repeat
this procedure, until termination. This `unranking' construction enforces that the $v_k$ yield a
strictly falling sequence $v_K > v_{K-1} > ... > v_1 \geq 0$ forming the $K$-combinations
$(v_K, v_{K-1}, ... \;, v_1)\equiv \VEC{v}$, which, by construction, are always ordered by size.

The reverse procedure is called the `ranking' of a $K$-combination and is encapsulated by the
function
\begin{align}
  \label{eq:Kcombination}
  V(\VEC{v}) = \binom{v_K}K+\ldots+\binom{v_2}2+\binom{v_1}1 \; .
\end{align}
We use the convention $\binom{v_i}i = 0$ if $v_i < i$, in other words, the
$K$-combination $ \VEC{v} = (K-1, K-2, ... \;, 0)$ represents $V=0$, a starting point of sorts.

It is known that for a given $K$ the $K$-combination of $V$ is unique, that ranking and unranking
are inverses and that they are independent of the size of $V$. Together, they form a bijection
on all of~$\mathbb{N}_0$.

\subsection{Skolem Polynomials for physicists\label{subsubsec:SkolemPolynomials}}

For our purposes it hinders us having to obey the constrictions of $K$-combination ordering into
falling sequences, instead, we want to employ unrestricted $K$-tuples
$(n_1, n_2, \;... \;, n_K) \equiv \VEC{n}$, whose entries are freely drawn from~$\mathbb{N}_0$,
since they are to represent the Fock excitations of $K$ independent modes~$j$.

A Skolem polynomial~\cite{rosser_review_1937} $\mathcal{S}$ of order $K$ is based on the
$K$-combination~(\ref{eq:Kcombination}), it has the form
\begin{subequations}
  \label{eq:DefSkolemPolynomial}
  \begin{align}
      \mathcal{S} (n_1, n_2, & \;... \;, n_K) \notag \\
  \label{eq:DefSkolemPolynoms1}
                  &  = \sum_{k=1}^{K} \left( n_1 + \ldots + n_k + k - 1 \atop k \right) \\
  \label{eq:DefSkolemPolynoms2}
                  &= \sum_{k=2}^{K} \left( n_1 + \ldots + n_k + k - 1 \atop k \right) + n_1.
\end{align}
\end{subequations}
The $\mathcal{S}$ inherit all uniqueness and invertibility features of the $K$-combinations whilst
managing to off-load the require\-ments for ordering into their construction, allowing us to use any
$K$-tuple $ \VEC{n}$ freely drawn from~$\mathbb{N}_0$.

Skolem polynomials~(\ref{eq:DefSkolemPolynoms2}) obviously fulfill the ``immediate-neighbors''
recurrence relations~(\ref{eq:recurrence})
\begin{align}\label{eq:RecurrencePropertySkolem}
	\mathcal{S} (n_1 \pm 1, n_2 \mp 1, ... \;, n_K) = \mathcal{S} (n_1, n_2, ... \;, n_K)  \pm 1 \; .
\end{align}

Skolem polynomials of fixed order also obey the `total-excitation island'--property: all $K$-tuples
$\VEC{n}\left.\right|_{N_{}}$ whose entries sum up to the same (total excitation) value,
$N_{} (\VEC{n}) \equiv \sum_{k=1}^K
n_k$, get mapped onto an uninterrupted
sequence~$\{z\} = \{ \mathcal{S} (\VEC{n}\left.\right|_{N_{}}) \}$ in~$\mathbb{N}_0$.

Additionally, Skolem polynomials of fixed order~$K$ obey ``diagonality''~\cite{morales1996enlarged,
  morales1999family}: the total-excitation islands themselves are ordered according to size: namely,
with $\sum_{k=1}^K x_k \equiv N_{} (\VEC{x}) > N_{} (\VEC{n})$,
\begin{align}
  \label{eq:Diagonality}
  \mathcal{S} (\VEC{n})  < \mathcal{S} (\VEC{x}) \quad \Longleftrightarrow \quad N_{} (\VEC{n}) < N_{} (\VEC{x}) \; .
\end{align}

\section{{{Application: Tridiagonal rendering}} of $\mathbf{H}_{\rm BH}$ \label{subsec:BHtridiagonalGlobal}}

With the bijection property inherited from $K$-combinations, Skolem polynomials of order~$K=3$ map the
computational bases of the pro\-duct Fock space
$\{\ket{\VEC{n}}\}_{n_1, n_2, n_3 = 0}^{N_1, N_2, N_3}$ onto $\{\|{z}\rrangle\}_{z = 0}^{D}$
without leaving gaps.

\subsection{ Tridiagonal rendering of  $\mathbf{H}_{1,2}$ \label{subsubsec:H12}}

Temporarily dropping reference to mode~`3' again (set $n_3 = $const.), it is straightforward to
check that using Skolem polynomials of order~$3$ leads to the formation of the dia\-go\-nal of
$\mathbf{H}_{\rm 1,2}$ such that within a total-excitation island, because of the
immediate-neighbors relation~(\ref{eq:RecurrencePropertySkolem}), the entry for $\ket{n_1-1, n_2+1}$
directly precedes the entry for $\ket{n_1, n_2}$ which is directly succeeded by the one for
$\ket{n_1+1, n_2-1}$. Total-excitation islands in turn, because of
dia\-go\-nality~(\ref{eq:Diagonality}), are sequentially aligned upward the dia\-go\-nal of
$\mathbf{H}_{\rm 1,2}$ as their total excitation value~$ N_{}$ increases (in steps of one).

Because of the bijection property, all this happens without gaps.

The off-dia\-go\-nal hopping terms for $j=1$ of Eq.~(\ref{Eq:BHHamiltonianHop}), as discussed in
Subsection~\ref{subsec:BHtridiagonalPreliminary}, conform with the order given to the dia\-go\-nal
and are rendered in the correct tri\-dia\-go\-nal form if the coupling
terms~$J_1 \sqrt{(n_1+1)n_2}$) are written into the matrix entries immediately below and immediately
to the right of that for $\ket{n_1, n_2}$.

\subsection{Tridiagonal rendering of 2-site Bose-Hubbard hamiltonians\label{subsubsec:2siteBHH}}

We emphasize that subsection~\ref{subsubsec:H12} shows how to tri\-dia\-go\-nalize a 2-site Bose-Hubbard
system, that is a hamiltonian~(\ref{eq:BHH}), which has cut-offs `2' for $j$ in
Eq.~(\ref{Eq:BHHamiltonianOnsite}) and `1' for $j$ in Eq.~(\ref{Eq:BHHamiltonianHop}).

For the tri\-dia\-go\-nali\-zation of the full hamiltonian~(\ref{eq:BHH}) we can proceed as in
subsection~\ref{subsubsec:H12} by ignoring, or setting constant mode $j = 1$, whilst taking care not
to count the dia\-go\-nal terms for $n_2$ in the onsite terms Eq.~(\ref{Eq:BHHamiltonianOnsite}) twice.

\subsection{Tridiagonal rendering of  $\mathbf{H}_{2,3}$ \label{subsubsec:H23}}

One can approach tri\-dia\-go\-nal rendering of $\mathbf{H}_{2,3}$ just like that of $\mathbf{H}_{1,2}$ in
Section~\ref{subsubsec:H12} with suitable relabeling of the modes.

The rendered matrix $\mathbf{H}_{2,3}$ requires a basis switch with respect to the basis of matrix
$\mathbf{H}_{1,2}$, which should be implemented as index permutations by a perfect shuffle matrix
$\mathbf{S}$, as discussed in Section~\ref{subsec:driveOMOhamiltonian}.

\begin{figure}[t]
    \centering
    \includegraphics[width=0.95\linewidth]{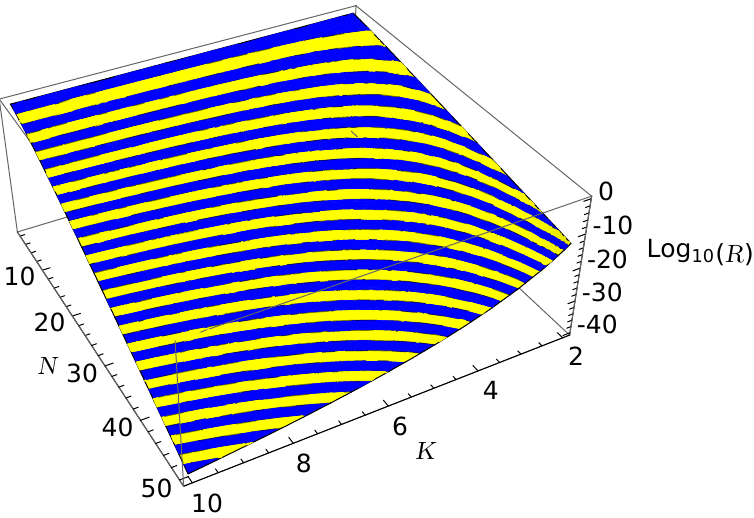}
    \caption{The ratio $R = {\cal N}_{N}^{K} / N^K $, see 
    Eq.~(\ref{eq:NumberInIsland}), quantifies the size of 
    an excitation island as compared to the na\"ively chosen size $N^K$
    of a hilbert space for $N$ excitations across $K$ sites. $R$ be\-comes 
    relatively very small for moderate values of $K$ and $N$.}
    \label{fig:shrinkRation}
\end{figure}

\subsection{Imposing symmetries on $\mathbf{H}_{\rm
    BH}~(\ref{eq:BHH})$ \label{subsubsec:HBHsymmetries}}

The hamiltonian~(\ref{eq:BHH}) conserves blocks with constant total excitation number
$N$ (see Sect.~\ref{subsubsec:SkolemPolynomials}
and Fig.~\ref{fig:shrinkRation}),
since its number-commutator vanishes.

The number of states (within such an island) with the same total excitation number~$N$ is~\cite{raventos2017cold}
\begin{align}
  \label{eq:NumberInIsland}
  {\cal N}_{N}^{K} = \binom{N+K-1}{N}
  = \frac{(N + K - 1)!}{N! \; (K-1)!} \; .
\end{align}
(Think of having to choose $N$ indistinguishable particles separated by $K-1$ indistinguishable
partition walls.)
This size, ${\cal N}$, of an excitation island grows much less than the na\"ively chosen size,
$N^K$, of a $K$-mode hilbert space with $N$ excitations,
constraining modelling to one total-excitation island is advised, see Fig.~\ref{fig:shrinkRation}.

Additionally, frequently, it is interesting to study degenerate systems with higher symmetries by
setting the chemical potentials $\mu_j = \mu$ equal and do the same thing for the intra-site
repulsion strengths~$U_j = U$ and the hopping matrix elements~ $J_{j} = J$.  In this case, matrix
$\mathbf{H}_{2,3}$, up to the required basis switch~(\ref{eq:Coupling_H_recycle}), is
\emph{identical} to $\mathbf{H}_{1,2}$, the code for one can be recycled to implement the other.

\subsection{Putting it all together: rendering of  $\mathbf{H}_{\rm BH}$ \label{subsubsec:HBH}}

Eq.~(\ref{eq:NumberInIsland}) quantifies the number $ {\cal N}_{N_{}}^{K}$ of basis elements on
islands with a total-excitation number~$N_{}$. For our number-conserving systems, the dynamics does
not leave such an island, if the entire island is correctly represented, without any accidental
cutoff.

Avoidance of such cutoffs is not a trivial proposition~\cite{fehske_exact_2008, zhang2010exact,raventos2017cold}, in
our approach, however, this is automatically taken care of by using the Skolem polynomials.

As an example we list the ${\cal N}_{6}^{3} =28 $ bases for $N=6$ excitations across $K=3$ sites  \\
\noindent
$N_3\!\!=$[0 0 1 0 1 2 0 1 2 3 0 1 2 3 4 0 1 2 3 4 5 0 1 2 3 4 5 6]
\\
\noindent
$N_2\!\!=$[0 1 0 2 1 0 3 2 1 0 4 3 2 1 0 5 4 3 2 1 0 6 5 4 3 2 1 0]
\\
\noindent
$N_1\!\!=$[6 5 5 4 4 4 3 3 3 3 2 2 2 2 2 1 1 1 1 1 1 0 0 0 0 0 0 0]

More generally, for a 3-site system, the state vectors $\ket{ 0, 0, N_{}}$ 
and $\ket{ N_{}, 0, 0}$ at the `extreme ends' of an island's pro\-duct hilbert space, have the 
$z$-entries~(\ref{eq:DefSkolemPolynomial}) [$m_N = \mathcal{S} (0, 0, N)$] and
[$M_N = \mathcal{S} (N, 0, 0)$] for lowest and highest value within the island, obeying 
$m_N + {\cal N}_{N}^{3} = M_N+1$.

The associated computational basis should therefore not range from $z = 0$ to $D$ but be restricted
to the intra-island form $\{\|{z}\rrangle\}_{z = m_N}^{M_N}$, compare
Sect.~\ref{subsubsec:SkolemPolynomials} and Fig.~\ref{fig:shrinkRation}.

Restriction to a total-excitation island also implies that the associated pro\-duct Fock space basis
should, instead of the unsuitable general sector
$\{\ket{\VEC{n}}\}_{n_1, n_2, n_3 = 0}^{N_1, N_2, N_3}$, use the exact cover of the total-excitation
island $\{\ket{\VEC{n}}\}_{\mathcal{S}^{-1} (m_N)}^{\mathcal{S}^{-1} (M_N)}$. Here,
$\mathcal{S}^{-1} (z)$ is the inverse of $\mathcal{S}(\VEC{n})$ and given by the unranking
construction yielding $\VEC{n}$'s $K$-combination $\VEC{v}$ of~Eq.~(\ref{eq:Kcombination}),
$\VEC{v}$ is then mapped to $\VEC{n}$ by suitable index inversions and subtractions:
$n_j = v_{K-j-1} - j - \sum_{k<j} v_k$.

\begin{widetext}
\section{Some Numerical Results\label{sec:NumericalSpeeds}}
\begin{figure}[h]
    \centering
    \includegraphics[width=0.32\linewidth,height=4.2cm]{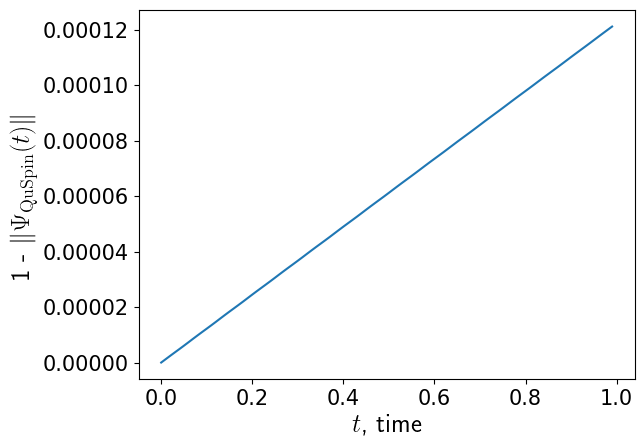}
    \includegraphics[width=0.32\linewidth,height=4.2cm]{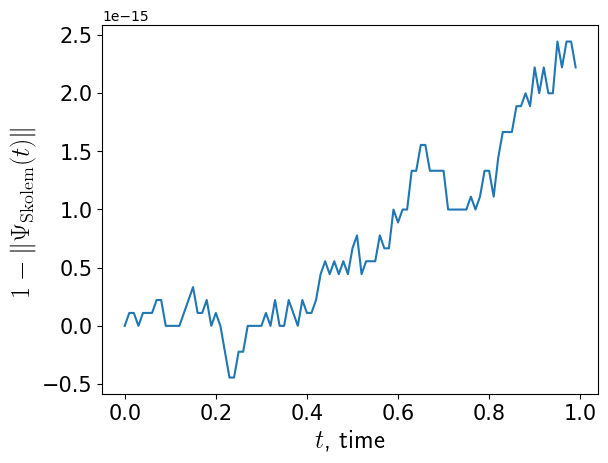}
    \includegraphics[width=0.32\linewidth,height=4.2cm]{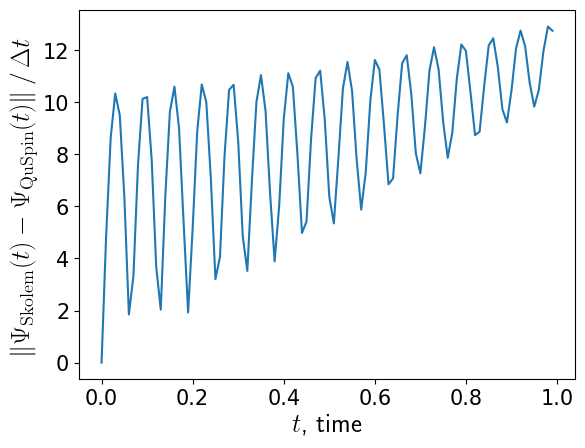}
    \caption{For a Bose-Hubbard system~\cite{QuSpinBHM}, with $U = 1$, $J = 1,$
      $\mu = 0,$ for $ N = 100$ bosons on $ K = 4$ sites ($\psi(t=0)=|100,0,0,0\rangle$), and time-steps $ dt = 0.01$, 
      the plots demonstrate that using QuSpin-software the size of the state, in the Euclidian $l^2$-norm, grows
      linearly with time (away from unity). Instead, our Skolem-approach uses a
      symplectic integrator preserving norm at machine 
      precision (note: scale on ordinate axis multiplied by $10^{15}$). 
      Also, for this example, our code runs roughly ten times faster~\cite{Codes}.}
    \label{fig:Norm_t}
\end{figure}
\end{widetext}

\section{Conclusion\label{sec:Conclusion}}

Our approach allows for a straight\-forward, transparent, flexible and low-resource
implementation of the exact tri\-dia\-go\-nali\-zation of standard opto\-mecha\-nical and Bose-Hubbard
models.

We highlight that implementation as unitary exponential propagators provides stability and flexibility, 
including second order precision in evolution time-steps, $\smallTimeStep$, at only linear cost.
Specifically, basis changes are implemented via index permutations, at marginal cost.

We emphasized that parts of our approach lend themselves to obvious code-recycling~\cite{Codes}.

\emph{Code availability:} All code used in this study can be found in~\cite{Codes}.

\textit{Acknowledgments.} 
D.I.B. was supported by Army Research Office (ARO) (grant W911NF-23-1-0288; program manager Dr.~James Joseph). The views and conclusions contained in this document are those of the authors and should not be interpreted as representing the official policies, either expressed or implied, of ARO, NSF, or the U.S. Government. The U.S. Government is authorized to reproduce and distribute reprints for Government purposes notwithstanding any copyright notation herein.

\bibliography{bibliography}

\begin{thebibliography}{21}%
\makeatletter
\providecommand \@ifxundefined [1]{%
 \@ifx{#1\undefined}
}%
\providecommand \@ifnum [1]{%
 \ifnum #1\expandafter \@firstoftwo
 \else \expandafter \@secondoftwo
 \fi
}%
\providecommand \@ifx [1]{%
 \ifx #1\expandafter \@firstoftwo
 \else \expandafter \@secondoftwo
 \fi
}%
\providecommand \natexlab [1]{#1}%
\providecommand \enquote  [1]{``#1''}%
\providecommand \bibnamefont  [1]{#1}%
\providecommand \bibfnamefont [1]{#1}%
\providecommand \citenamefont [1]{#1}%
\providecommand \href@noop [0]{\@secondoftwo}%
\providecommand \href [0]{\begingroup \@sanitize@url \@href}%
\providecommand \@href[1]{\@@startlink{#1}\@@href}%
\providecommand \@@href[1]{\endgroup#1\@@endlink}%
\providecommand \@sanitize@url [0]{\catcode `\\12\catcode `\$12\catcode
  `\&12\catcode `\#12\catcode `\^12\catcode `\_12\catcode `\%12\relax}%
\providecommand \@@startlink[1]{}%
\providecommand \@@endlink[0]{}%
\providecommand \url  [0]{\begingroup\@sanitize@url \@url }%
\providecommand \@url [1]{\endgroup\@href {#1}{\urlprefix }}%
\providecommand \urlprefix  [0]{URL }%
\providecommand \Eprint [0]{\href }%
\providecommand \doibase [0]{https://doi.org/}%
\providecommand \selectlanguage [0]{\@gobble}%
\providecommand \bibinfo  [0]{\@secondoftwo}%
\providecommand \bibfield  [0]{\@secondoftwo}%
\providecommand \translation [1]{[#1]}%
\providecommand \BibitemOpen [0]{}%
\providecommand \bibitemStop [0]{}%
\providecommand \bibitemNoStop [0]{.\EOS\space}%
\providecommand \EOS [0]{\spacefactor3000\relax}%
\providecommand \BibitemShut  [1]{\csname bibitem#1\endcsname}%
\let\auto@bib@innerbib\@empty
\bibitem [{\citenamefont {Coakley}\ and\ \citenamefont
  {Rokhlin}(2012)}]{Coakley_DlnD_12}%
  \BibitemOpen
  \bibfield  {author} {\bibinfo {author} {\bibfnamefont {E.}~\bibnamefont
  {Coakley}}\ and\ \bibinfo {author} {\bibfnamefont {V.}~\bibnamefont
  {Rokhlin}},\ }\bibfield  {title} {\bibinfo {title} {A fast divide-and-conquer
  algorithm for computing the spectra of real symmetric tridiagonal matrices},\
  }\href {https://doi.org/10.1016/j.acha.2012.06.003} {\bibfield  {journal}
  {\bibinfo  {journal} {Appl. Comp. Harm. Anal.}\ ,\ \bibinfo {pages}
  {379–414}} (\bibinfo {year} {2012})}\BibitemShut {NoStop}%
\bibitem [{Dia()}]{Diagonalize3DiagFootnote}%
  \BibitemOpen
  \href@noop {} {}\bibinfo {note} {We are not aware of a standard library
  implementing method~\protect{\cite{Coakley_DlnD_12}} and did not implement it
  ourselves. All numerical results reported here are based on readily available
  libraries (\texttt{scipy.linalg.eigvalsh\_tridiagonal}), these allow for a
  linear speedup of the diagonlization of three-triagonal matrices of size $D
  \times D$ using ${\cal{O}}(D^2)$ steps, instead of ${\cal{O}}(D^3)$ steps for
  general matrices.}\BibitemShut {Stop}%
\bibitem [{\citenamefont {Yoshida}(1993)}]{Yoshida__Proc93}%
  \BibitemOpen
  \bibfield  {author} {\bibinfo {author} {\bibfnamefont {H.}~\bibnamefont
  {Yoshida}},\ }\bibfield  {title} {\bibinfo {title} {Recent progress in the
  theory and application of symplectic integrators},\ }in\ \href
  {https://adsabs.harvard.edu/full/1993CeMDA..56...27Y} {\emph {\bibinfo
  {booktitle} {Qualitative and Quantitative Behaviour of Planetary Systems:
  Proceedings of the Third Alexander von Humboldt Colloquium on Celestial
  Mechanics}}}\ (\bibinfo {organization} {Springer},\ \bibinfo {year} {1993})\
  pp.\ \bibinfo {pages} {27--43}\BibitemShut {NoStop}%
\bibitem [{\citenamefont {Chin}(2009)}]{Chin_PRE07}%
  \BibitemOpen
  \bibfield  {author} {\bibinfo {author} {\bibfnamefont {S.~A.}\ \bibnamefont
  {Chin}},\ }\bibfield  {title} {\bibinfo {title} {Explicit symplectic
  integrators for solving nonseparable hamiltonians},\ }\href
  {https://doi.org/10.1103/PhysRevE.80.037701} {\bibfield  {journal} {\bibinfo
  {journal} {Phys. Rev. E}\ }\textbf {\bibinfo {volume} {80}},\ \bibinfo
  {pages} {037701} (\bibinfo {year} {2009})}\BibitemShut {NoStop}%
\bibitem [{\citenamefont {{\'C}iri{\'c}}\ \emph {et~al.}(2023)\citenamefont
  {{\'C}iri{\'c}}, \citenamefont {Bondar},\ and\ \citenamefont
  {Steuernagel}}]{Ciric_EJPP23}%
  \BibitemOpen
  \bibfield  {author} {\bibinfo {author} {\bibfnamefont {M.}~\bibnamefont
  {{\'C}iri{\'c}}}, \bibinfo {author} {\bibfnamefont {D.~I.}\ \bibnamefont
  {Bondar}},\ and\ \bibinfo {author} {\bibfnamefont {O.}~\bibnamefont
  {Steuernagel}},\ }\bibfield  {title} {\bibinfo {title} {Exponential unitary
  integrators for nonseparable quantum hamiltonians},\ }\href
  {https://doi.org/10.1140/epjp/s13360-023-03819-3} {\bibfield  {journal}
  {\bibinfo  {journal} {Eur. Phys. J. Plus}\ }\textbf {\bibinfo {volume}
  {138}},\ \bibinfo {pages} {1} (\bibinfo {year} {2023})},\ \Eprint
  {https://arxiv.org/abs/2211.08155} {2211.08155} \BibitemShut {NoStop}%
\bibitem [{\citenamefont {Cabrera}\ \emph {et~al.}(2015)\citenamefont
  {Cabrera}, \citenamefont {Bondar}, \citenamefont {Jacobs},\ and\
  \citenamefont {Rabitz}}]{Cabrera_PRA15}%
  \BibitemOpen
  \bibfield  {author} {\bibinfo {author} {\bibfnamefont {R.}~\bibnamefont
  {Cabrera}}, \bibinfo {author} {\bibfnamefont {D.~I.}\ \bibnamefont {Bondar}},
  \bibinfo {author} {\bibfnamefont {K.}~\bibnamefont {Jacobs}},\ and\ \bibinfo
  {author} {\bibfnamefont {H.~A.}\ \bibnamefont {Rabitz}},\ }\bibfield  {title}
  {\bibinfo {title} {{Efficient method to generate time evolution of the Wigner
  function for open quantum systems}},\ }\href
  {https://doi.org/10.1103/PhysRevA.92.042122} {\bibfield  {journal} {\bibinfo
  {journal} {Phys. Rev. A}\ }\textbf {\bibinfo {volume} {92}},\ \bibinfo
  {pages} {042122} (\bibinfo {year} {2015})},\ \Eprint
  {https://arxiv.org/abs/1212.3406} {1212.3406} \BibitemShut {NoStop}%
\bibitem [{\citenamefont {Javanainen}\ and\ \citenamefont
  {Ruostekoski}(2006)}]{Javanainen_JPA06}%
  \BibitemOpen
  \bibfield  {author} {\bibinfo {author} {\bibfnamefont {J.}~\bibnamefont
  {Javanainen}}\ and\ \bibinfo {author} {\bibfnamefont {J.}~\bibnamefont
  {Ruostekoski}},\ }\bibfield  {title} {\bibinfo {title} {Symbolic calculation
  in development of algorithms: split-step methods for the gross--pitaevskii
  equation},\ }\href {https://doi.org/10.1088/0305-4470/39/12/L02} {\bibfield
  {journal} {\bibinfo  {journal} {J. Phys. A: Math. Gen.}\ }\textbf {\bibinfo
  {volume} {39}},\ \bibinfo {pages} {L179} (\bibinfo {year}
  {2006})}\BibitemShut {NoStop}%
\bibitem [{\citenamefont {Weiße}\ and\ \citenamefont
  {Fehske}(2008)}]{fehske_exact_2008}%
  \BibitemOpen
  \bibfield  {author} {\bibinfo {author} {\bibfnamefont {A.}~\bibnamefont
  {Weiße}}\ and\ \bibinfo {author} {\bibfnamefont {H.}~\bibnamefont
  {Fehske}},\ }\bibfield  {title} {\bibinfo {title} {Exact {Diagonalization}
  {Techniques}},\ }in\ \href {https://doi.org/10.1007/978-3-540-74686-7_18}
  {\emph {\bibinfo {booktitle} {Computational {Many}-{Particle} {Physics}}}},\
  Vol.\ \bibinfo {volume} {739},\ \bibinfo {editor} {edited by\ \bibinfo
  {editor} {\bibfnamefont {H.}~\bibnamefont {Fehske}}, \bibinfo {editor}
  {\bibfnamefont {R.}~\bibnamefont {Schneider}},\ and\ \bibinfo {editor}
  {\bibfnamefont {A.}~\bibnamefont {Weiße}}}\ (\bibinfo  {publisher} {Springer
  Berlin Heidelberg},\ \bibinfo {address} {Berlin, Heidelberg},\ \bibinfo
  {year} {2008})\ pp.\ \bibinfo {pages} {529--544}\BibitemShut {NoStop}%
\bibitem [{LAP()}]{LAPACK}%
  \BibitemOpen
  \href {https://netlib.org/lapack/} {\bibinfo {title} {{\it LAPACK -- Linear
  Algebra PACKage}}}\BibitemShut {NoStop}%
\bibitem [{\citenamefont {Mancini}\ \emph {et~al.}(1997)\citenamefont
  {Mancini}, \citenamefont {Man'ko},\ and\ \citenamefont
  {Tombesi}}]{Mancini__PRA97}%
  \BibitemOpen
  \bibfield  {author} {\bibinfo {author} {\bibfnamefont {S.}~\bibnamefont
  {Mancini}}, \bibinfo {author} {\bibfnamefont {V.~I.}\ \bibnamefont
  {Man'ko}},\ and\ \bibinfo {author} {\bibfnamefont {P.}~\bibnamefont
  {Tombesi}},\ }\bibfield  {title} {\bibinfo {title} {Ponderomotive control of
  quantum macroscopic coherence},\ }\href
  {https://doi.org/10.1103/PhysRevA.55.3042} {\bibfield  {journal} {\bibinfo
  {journal} {Phys. Rev. A}\ }\textbf {\bibinfo {volume} {55}},\ \bibinfo
  {pages} {3042} (\bibinfo {year} {1997})}\BibitemShut {NoStop}%
\bibitem [{\citenamefont {Kim}\ \emph {et~al.}(2002)\citenamefont {Kim},
  \citenamefont {Son}, \citenamefont {Bu\ifmmode~\check{z}\else \v{z}\fi{}ek},\
  and\ \citenamefont {Knight}}]{Kim__PRA02}%
  \BibitemOpen
  \bibfield  {author} {\bibinfo {author} {\bibfnamefont {M.~S.}\ \bibnamefont
  {Kim}}, \bibinfo {author} {\bibfnamefont {W.}~\bibnamefont {Son}}, \bibinfo
  {author} {\bibfnamefont {V.}~\bibnamefont {Bu\ifmmode~\check{z}\else
  \v{z}\fi{}ek}},\ and\ \bibinfo {author} {\bibfnamefont {P.~L.}\ \bibnamefont
  {Knight}},\ }\bibfield  {title} {\bibinfo {title} {Entanglement by a beam
  splitter: Nonclassicality as a prerequisite for entanglement},\ }\href
  {https://doi.org/10.1103/PhysRevA.65.032323} {\bibfield  {journal} {\bibinfo
  {journal} {Phys. Rev. A}\ }\textbf {\bibinfo {volume} {65}},\ \bibinfo
  {pages} {032323} (\bibinfo {year} {2002})},\ \Eprint
  {https://arxiv.org/abs/quant-ph/0106136} {quant-ph/0106136} \BibitemShut
  {NoStop}%
\bibitem [{\citenamefont {Qvarfort}\ \emph {et~al.}(2019)\citenamefont
  {Qvarfort}, \citenamefont {Serafini}, \citenamefont {Xuereb}, \citenamefont
  {R{\"a}tzel},\ and\ \citenamefont {Bruschi}}]{Qvarfort_NJP19}%
  \BibitemOpen
  \bibfield  {author} {\bibinfo {author} {\bibfnamefont {S.}~\bibnamefont
  {Qvarfort}}, \bibinfo {author} {\bibfnamefont {A.}~\bibnamefont {Serafini}},
  \bibinfo {author} {\bibfnamefont {A.}~\bibnamefont {Xuereb}}, \bibinfo
  {author} {\bibfnamefont {D.}~\bibnamefont {R{\"a}tzel}},\ and\ \bibinfo
  {author} {\bibfnamefont {D.~E.}\ \bibnamefont {Bruschi}},\ }\bibfield
  {title} {\bibinfo {title} {Enhanced continuous generation of non-gaussianity
  through optomechanical modulation},\ }\href
  {https://doi.org/10.1088/1367-2630/ab1b9e} {\bibfield  {journal} {\bibinfo
  {journal} {New J. Phys.}\ }\textbf {\bibinfo {volume} {21}},\ \bibinfo
  {pages} {055004} (\bibinfo {year} {2019})}\BibitemShut {NoStop}%
\bibitem [{\citenamefont {{Hatano}}\ and\ \citenamefont
  {{Suzuki}}(2005)}]{Hatano_LN05}%
  \BibitemOpen
  \bibfield  {author} {\bibinfo {author} {\bibfnamefont {N.}~\bibnamefont
  {{Hatano}}}\ and\ \bibinfo {author} {\bibfnamefont {M.}~\bibnamefont
  {{Suzuki}}},\ }\bibfield  {title} {\bibinfo {title} {{Finding Exponential
  Product Formulas of Higher Orders}},\ }in\ \href
  {https://doi.org/10.1007/11526216_2} {\emph {\bibinfo {booktitle} {Lecture
  Notes in Physics, Berlin Springer Verlag}}},\ Vol.\ \bibinfo {volume} {679},\
  \bibinfo {editor} {edited by\ \bibinfo {editor} {\bibfnamefont
  {A.}~\bibnamefont {{Das}}}\ and\ \bibinfo {editor} {\bibfnamefont {B.~K.}\
  \bibnamefont {{Chakrabarti}}}}\ (\bibinfo {year} {2005})\ p.~\bibinfo {pages}
  {37},\ \Eprint {https://arxiv.org/abs/math-ph/0506007} {math-ph/0506007}
  \BibitemShut {NoStop}%
\bibitem [{Bis()}]{BisectionFootnote}%
  \BibitemOpen
  \href@noop {} {}\bibinfo {note} {To find this largest value the use of a
  bisection method can provide a very considerable speedup.}\BibitemShut
  {Stop}%
\bibitem [{\citenamefont {Rosser}(1937)}]{rosser_review_1937}%
  \BibitemOpen
  \bibfield  {author} {\bibinfo {author} {\bibfnamefont {B.}~\bibnamefont
  {Rosser}},\ }\bibfield  {title} {\bibinfo {title} {Reviewed work: Über die
  zurückführbarkeit einiger durch rekursionen definierter relationen auf
  "arithmetische."},\ }\href {https://doi.org/10.2307/2267375} {\bibfield
  {journal} {\bibinfo  {journal} {The Journal of Symbolic Logic}\ }\textbf
  {\bibinfo {volume} {2}},\ \bibinfo {pages} {85} (\bibinfo {year}
  {1937})}\BibitemShut {NoStop}%
\bibitem [{\citenamefont {Morales}\ and\ \citenamefont
  {Lew}(1996)}]{morales1996enlarged}%
  \BibitemOpen
  \bibfield  {author} {\bibinfo {author} {\bibfnamefont {L.~B.}\ \bibnamefont
  {Morales}}\ and\ \bibinfo {author} {\bibfnamefont {J.~S.}\ \bibnamefont
  {Lew}},\ }\bibfield  {title} {\bibinfo {title} {An enlarged family of packing
  polynomials on multidimensional lattices},\ }\href
  {https://www.academia.edu/download/60432985/bf0120128120190829-7322-v1c8po.pdf}
  {\bibfield  {journal} {\bibinfo  {journal} {Math. Systems Theory}\ }\textbf
  {\bibinfo {volume} {29}},\ \bibinfo {pages} {293} (\bibinfo {year}
  {1996})}\BibitemShut {NoStop}%
\bibitem [{\citenamefont {Morales}\ and\ \citenamefont
  {Arredondo}(1999)}]{morales1999family}%
  \BibitemOpen
  \bibfield  {author} {\bibinfo {author} {\bibfnamefont {L.~B.}\ \bibnamefont
  {Morales}}\ and\ \bibinfo {author} {\bibfnamefont {J.~H.}\ \bibnamefont
  {Arredondo}},\ }\bibfield  {title} {\bibinfo {title} {A family of
  asymptotically e (n- 1)! polynomial orders of $n^n$},\ }\href
  {https://www.academia.edu/download/87897284/a_3A100632922420220220622-1-fud43h.pdf}
  {\bibfield  {journal} {\bibinfo  {journal} {Order}\ }\textbf {\bibinfo
  {volume} {16}},\ \bibinfo {pages} {195} (\bibinfo {year} {1999})}\BibitemShut
  {NoStop}%
\bibitem [{\citenamefont {Ravent{\'o}s}\ \emph {et~al.}(2017)\citenamefont
  {Ravent{\'o}s}, \citenamefont {Gra{\ss}}, \citenamefont {Lewenstein},\ and\
  \citenamefont {Juli{\'a}-D{\'\i}az}}]{raventos2017cold}%
  \BibitemOpen
  \bibfield  {author} {\bibinfo {author} {\bibfnamefont {D.}~\bibnamefont
  {Ravent{\'o}s}}, \bibinfo {author} {\bibfnamefont {T.}~\bibnamefont
  {Gra{\ss}}}, \bibinfo {author} {\bibfnamefont {M.}~\bibnamefont
  {Lewenstein}},\ and\ \bibinfo {author} {\bibfnamefont {B.}~\bibnamefont
  {Juli{\'a}-D{\'\i}az}},\ }\bibfield  {title} {\bibinfo {title} {Cold bosons
  in optical lattices: a tutorial for exact diagonalization},\ }\href
  {https://doi.org/10.1088/1361-6455/aa68b1} {\bibfield  {journal} {\bibinfo
  {journal} {J. Phys. B}\ }\textbf {\bibinfo {volume} {50}},\ \bibinfo {pages}
  {113001} (\bibinfo {year} {2017})}\BibitemShut {NoStop}%
\bibitem [{\citenamefont {Zhang}\ and\ \citenamefont
  {Dong}(2010)}]{zhang2010exact}%
  \BibitemOpen
  \bibfield  {author} {\bibinfo {author} {\bibfnamefont {J.}~\bibnamefont
  {Zhang}}\ and\ \bibinfo {author} {\bibfnamefont {R.}~\bibnamefont {Dong}},\
  }\bibfield  {title} {\bibinfo {title} {Exact diagonalization: the
  bose--hubbard model as an example},\ }\href
  {https://doi.org/10.1088/0143-0807/31/3/016} {\bibfield  {journal} {\bibinfo
  {journal} {Eur. J. Phys.}\ }\textbf {\bibinfo {volume} {31}},\ \bibinfo
  {pages} {591} (\bibinfo {year} {2010})}\BibitemShut {NoStop}%
\bibitem [{QuS(2019)}]{QuSpinBHM}%
  \BibitemOpen
  \href
  {https://github.com/weinbe58/QuSpin/blob/master/examples/notebooks/BHM.ipynb}
  {\bibinfo {title} {Coding the bose-hubbard hamiltonian with quspin}},\
  \bibinfo {howpublished}
  {\href{https://github.com/weinbe58/QuSpin/blob/master/examples/notebooks/BHM.ipynb}
  {GitHub Link}} (\bibinfo {year} {2019})\BibitemShut {NoStop}%
\bibitem [{Cod(2024)}]{Codes}%
  \BibitemOpen
  \href
  {https://github.com/dibondar/SkolemPropagatorBoseHubbard/blob/master/demo_second_ord_skolem_propagator.ipynb}
  {\bibinfo {title} {Symplectic split-operator propagators from tridiagonalized
  multi-mode bosonic hilbert spaces for bose-hubbard}},\ \bibinfo
  {howpublished}
  {\href{https://github.com/dibondar/SkolemPropagatorBoseHubbard/blob/master/demo_second_ord_skolem_propagator.ipynb}
  {GitHub Link}} (\bibinfo {year} {2024})\BibitemShut {NoStop}%
\end{thebibliography}%


\end{document}